\documentclass[reprint,secnumarabic,amssymb, nobibnotes, aps, pre,groupedaddress,superscriptaddress]{revtex4-1}
\usepackage{graphicx}
\usepackage{dcolumn}
\usepackage{bm}
\usepackage{siunitx}
\usepackage[dvipsnames]{xcolor}
\usepackage{lipsum}
\usepackage{verbatim}
\usepackage{soul}
\usepackage{comment}
\usepackage{amsmath} 
\usepackage{amssymb} 

\renewcommand\thesubsection{\Alph{subsection}}

\begin{document}


\title{Positron acceleration in plasma waves driven by non-neutral fireball beams}

	\author{Thales Silva}%
	\email{thales.silva@tecnico.ulisboa.pt}
	\affiliation{GoLP/Instituto de Plasmas e Fus\~ao Nuclear, Instituto Superior T\'ecnico, Universidade de Lisboa, 1049-001 Lisbon, Portugal}
	\author{Jorge Vieira}
	\email{jorge.vieira@tecnico.ulisboa.pt}
	\affiliation{GoLP/Instituto de Plasmas e Fus\~ao Nuclear, Instituto Superior T\'ecnico, Universidade de Lisboa, 1049-001 Lisbon, Portugal}
	\date{\today}%

\begin{abstract}
Plasma-based positron acceleration is still an open question, as the most efficient regimes for electron acceleration (quasi-linear and blowout) are not directly applicable to positrons. Nevertheless, positron acceleration is a stepping stone on the path toward a plasma-based lepton collider. In this work, we propose a scheme for positron acceleration based on the spatial overlap of a driver (electron or laser) beam and a positron beam, also known as a fireball beam. Under appropriate conditions, these beams can self-consistently evolve toward a hollow driver and focused positron beam on-axis, driving plasma waves suitable for positron acceleration. This evolution seems to be a manifestation of the current filamentation instability. We discuss how the self-consistent dynamics affect the beam quality and perform a preliminary tolerance study.
\end{abstract}

\maketitle

\section{Introduction}
Plasma accelerators are promising candidates for the next generations of particle colliders \cite{2021Albert}. The breakdown of the accelerator components limits the electromagnetic fields supported by conventional RF technology. In contrast, plasmas, being an ionized medium, can sustain electric fields that can be orders of magnitude higher than conventional machines \cite{2012Malka}. In theory, this allows for more compact machines and facilities \cite{2020Assmann}.

The quality of the accelerated electron beams has improved substantially over the last decade. Thus, plasma-based accelerators now routinely deliver relativistic electron and x-ray beams for high-energy-density science \cite{2018Wood}, quantum electrodynamics \cite{2018Cole}, biology \cite{2015Cole}, and material science \cite{2016He}. Furthermore, as these accelerators began providing sub-percent energy spread electron bunches, pioneering experiments demonstrated lasing in a free-electron laser driven by electron beams from plasma accelerators \cite{2021Wang,2022Pompili,2022Labat}.

The ultimate goal of plasma accelerator research in the context of high energy physics is to project an electron-positron plasma-based collider \cite{2019Cros,2021Albert}. Here, demonstrating stable, high-quality positron acceleration is a key step \cite{2015Corde,2014Vieira,2015Jain,2019Diederichs,2021Silva,2021Zhou,2022Reichwein,2022Zhou,2021Hue}. Positron acceleration is not as straightforward as for electrons because the most common acceleration regimes, when the driver is sufficiently intense to drive nonlinear waves, are not proper for positron focusing. Due to the ion and electron mass discrepancy, the two plasma species react at different timescales; consequently, plasma waves tend to defocus positively charged particles almost everywhere if the driver is sufficiently intense. To circumvent this issue, one needs to shape the plasma wave to create focusing regions for positrons. One approach uses a long positron beam as the driver, leading to an energy transfer from the beam head to the tail \cite{2015Corde}. Other schemes rely on shaping the plasma wave driver \cite{2014Vieira,2015Jain}, the plasma itself \cite{2019Diederichs,2021Silva}, both \cite{2021Zhou,2022Reichwein,2022Zhou}, direct laser acceleration by ultra-intense lasers \cite{arxivMartinez}, or lower intensity drivers (linear waves) \cite{2021Hue}.

For example, Refs. \cite{2014Vieira,2015Jain} explored non-Gaussian drivers using lasers with orbital angular momentum or hollow electron beams, where the driver has null intensity or density on the symmetry axis. In these cases, the plasma waves driven present a dense electron filament on the axis, resulting in a focusing force for positrons. These contrast with usual regimes of plasma acceleration \cite{2006Lu,2007Lu}, where the driver profile is close to a Gaussian, with the peak density or intensity maximum at the symmetry axis; thus, the plasma electrons are completely repelled (either by the particle beam fields or the laser ponderomotive force) away from the axis, leaving a near-spherical ion cavity on the wake of the driver.

In addition to positron acceleration, the interplay between electron and positron beams propagating in plasmas has another potential in studying plasma micro-instabilities. These instabilities can play a crucial role in the astrophysical environment, which led to the proposition of many schemes to probe it in the laboratory \cite{2015Huntington,2017Nerush,2019Zhang,2020Silva,2020Raj,2020Shukla_PRR,2020Zhang,2022Shukla,2022Boella}. Among them is the propagation of neutral beams composed of spatial overlap of electrons and positrons \cite{2015Sarri,2018Shukla,2020Shukla,2021Arrowsmith}, sometimes referred to as fireball beams, in analogy with cosmic fireballs (c.f. \cite{1990Shemi}).

In this paper, we show that the propagation of fireball beams in plasmas can result in energy transfer from the electron to the positron beam under appropriate conditions. In particular, non-neutral fireball beams with Gaussian profiles may self-consistently evolve toward a hollow electron beam and a focused positron beam on-axis. Similar results are also observed by replacing the electron beam with a laser beam. Thus, using Gaussian beams, more readily available in laboratories, one can accelerate positrons using a configuration analogous to the proposed in Refs. \cite{2014Vieira,2015Jain}, relaxing the requirements to shape the driver. We use theory and particle-in-cell simulations to demonstrate the main features of the configuration. Additionally, we perform a preliminary tolerance study to understand how different beam parameters and imperfections can impact the configuration.

The paper is organized as follows: in Sec. \ref{sec:dyn}, we present the dynamics of non-neutral fireball beams in plasma and show how the self-consistent dynamics leads to a hollow electron beam circumventing a focused positron beam; in Sec. \ref{sec:imp}, we perform a tolerance study for imperfections and beam parameters that allow the positron acceleration process to be stable over long propagation distances; in Sec. \ref{sec:laser}, we show results for the analogous scheme when the driver is a laser beam; in Sec. \ref{sec:disc}, we briefly discuss our results and their consequences for positron acceleration and filamentation in the laboratory; in Sec. \ref{sec:concl}, we present our conclusions.

\section{Non-neutral fireball beam dynamics}
\label{sec:dyn}
Hollow electron or laser beams can drive plasma waves proper for positron acceleration \cite{2014Vieira,2015Jain}, characterized by an on-axis focusing force for positively charged particles. Here, we propose to combine an electron bunch or a laser pulse with a positron bunch to create such a hollow structure in the driver. We consider the spatial overlap of a Gaussian driver (electron or laser beam) and positron beams; we primarily focus on an electron beam driver, but we demonstrate the analogous configuration for a laser driver in Sec. \ref{sec:laser}.
\begin{figure}[t]
    \centering
    \includegraphics[width=.99\linewidth]{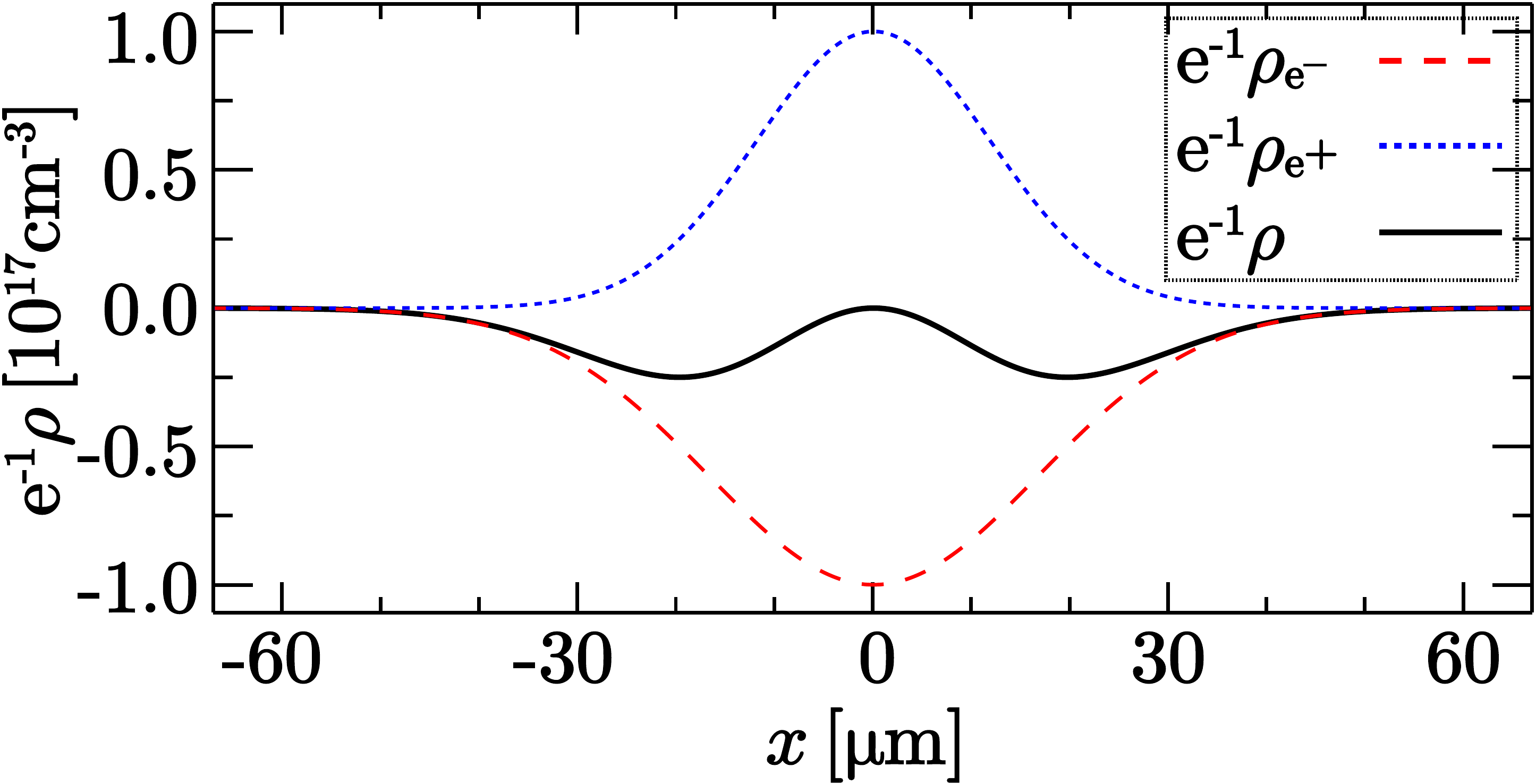}
    \caption{Transverse charge density at the beam center $z=0$ for the fireball (black, solid line), and the individual positron (blue, dotted) and electron beams (red, dashed).}
    \label{fig:beam_profile}
\end{figure}

Let
\begin{equation}
        \rho = \sum_\alpha q_{\alpha} n_{\alpha} \exp\left(-\frac{z^2}{2\sigma_{z,\alpha}^2}\right)\exp\left(-\frac{r^2}{2\sigma^2_{r,\alpha}}\right)
    \label{eq:dens_hollow}
\end{equation}
be the total charge density of all beam driver species (the index $\alpha$ sums over the beam species $e^{-}$ for electrons and $e^{+}$ for positrons), where $q_\alpha$ is the species charge, $n_\alpha$ is the peak density, $z$ and $r$ are the longitudinal and radial coordinates, and $\sigma_{z,\alpha}$ and $\sigma_{r,\alpha}$ are the beam size in the propagation and transverse directions, respectively. Assuming an electron and a positron beam such that the peak density and longitudinal sizes are equal ($n_{\mathrm{e}^-} = n_{\mathrm{e}^+} = n_b$, $\sigma_{z,\mathrm{e}^-} = \sigma_{z,\mathrm{e}^+}$), but the electron beam is larger than the positron radially ($\sigma_{r,\mathrm{e}^-}>\sigma_{r,\mathrm{e}^+}$), then $\rho = 0$ at $r = 0$ and $r \rightarrow \infty$, and $\rho$ has a minimum at $r = 2 \sigma_{r,\mathrm{e}^-}\sigma_{r,\mathrm{e}^+}[\log(\sigma_{r,\mathrm{e}^-}/\sigma_{r,\mathrm{e}^+})/(\sigma_{r,\mathrm{e}^-}^2-\sigma_{r,\mathrm{e}^+}^2)]^{1/2}$, effectively reproducing a hollow density profile as used in Ref. \cite{2015Jain}. We display an example in Fig. \ref{fig:beam_profile}, which shows the transverse profile of the individual positron (blue, dotted line) and electron (red, dashed line) beams and the resulting doughnut-like sum [Eq. \eqref{eq:dens_hollow}, black, solid line] at $z=0$ for a case where $\sigma_{r,\mathrm{e}^-} =  \SI{17}{\micro\meter}$ and $\sigma_{r,\mathrm{e}^+} = \SI{12}{\micro\meter}$, and $n_b = \SI{1e17}{\centi\meter^{-3}}$. These and $\sigma_{z,\mathrm{e}^-} = \sigma_{z,\mathrm{e}^+} = \SI{17}{\micro\meter}$ are the fiducial parameters that we use to demonstrate the  main features of the proposed configuration throughout the paper. For these parameters, the charge of the electrons and positron beam is $\SI{1.2}{\nano C}$ and $\SI{0.6}{\nano C}$, respectively. Note that simulations were performed with normalized units; thus, these numbers appear when we assume $n_0 = \SI{1e17}{\centi\meter^{-3}}$, to have beam parameters close to FACET-II \cite{2019Yakimenko}. In practice, our results are valid for a wide range of parameters.
\begin{figure}[b]
    \centering
    \includegraphics[width=.99\linewidth]{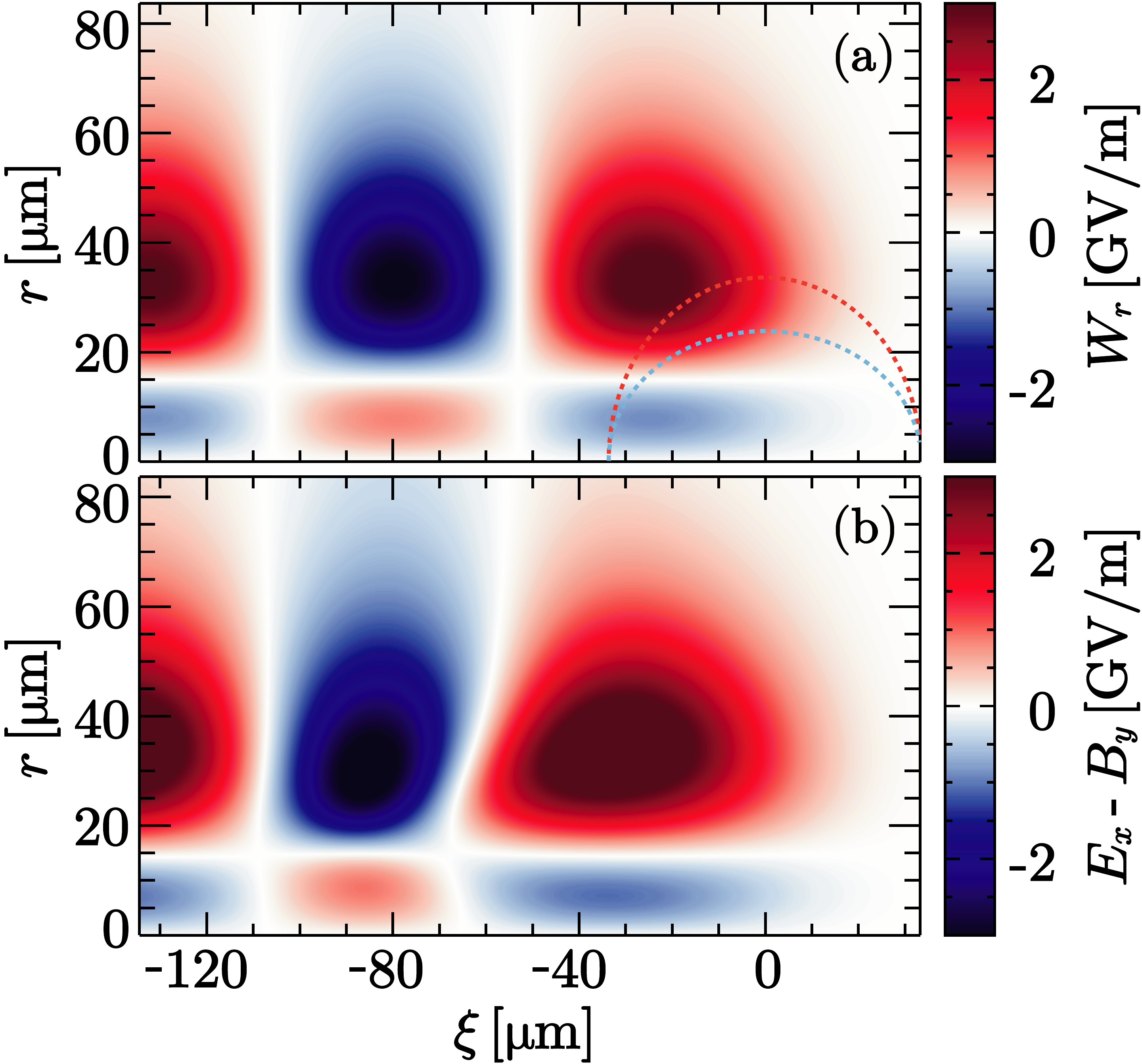}
    \caption{Transverse wakefield linear response to the fireball beam propagating in the plasma. (a) Theoretical result obtained from Eq. \eqref{eq:transv_wakefield}. The dashed lines are the contour of the electron (red, larger radius) and positron (blue, smaller radius) beams density at the $1/e^2$ of the peak value. (b) Simulation result ($E_x-B_y$) for the same parameters as in (a).}
    \label{fig:init_fields}
\end{figure}
\subsection{Linear regime}
We use linear wakefield theory \cite{1987Chen,1987Katsouleas,2016Muggli} to understand the linear plasma response to the propagation of the non-neutral fireball beam. Hereafter, we assume the plasma density ($n_0$) is the same as the individual peak beam density, i.e., $n_0 = n_b$. Assuming cylindrical symmetry and a separable charge density profile for the beams [Eq. \eqref{eq:dens_hollow}], $\rho = en_b\rho_r(r) \rho_z(\xi)$, the transverse wakefield is given by
\begin{equation}
    \frac{e W_r\left(\xi,r\right)}{m_e c \omega_p} = \frac{dR}{dr} \int_\xi^\infty \rho_z(\xi^\prime) \sin\left[k_p\left(\xi-\xi^\prime\right)\right] d\xi^\prime,
    \label{eq:transv_wakefield}
\end{equation}
where
\begin{multline*}
    R(r) = k_p^2 K_0(k_p r) \int_0^r r^\prime \rho_r(r^\prime) I_0(k_p r^\prime) dr^\prime \\+ k_p^2 I_0(k_p r) \int_r^\infty r^\prime \rho_r(r^\prime) K_0(k_p r^\prime) dr^\prime,
\end{multline*}
$K_0$ and $I_0$ are the zero-order modified Bessel functions of the first and second kind, $k_p = \omega_p/c$, $c$ is the speed of light in vacuum, $\omega_p = (4\pi n_0 e^2/m_e)^{1/2}$ is the electron plasma frequency, $e$ is the elementary charge, $m_e$ is the electron mass, $\xi = z - c t$ is the longitudinal coordinate that follows the beam ($\xi = 0$ is the beam center), and $t$ is the time. The transverse wakefield $W_r$ measures the transverse electromagnetic fields $W_r \equiv E_r - B_\theta$ acting on relativistic beams. Figure \ref{fig:init_fields}(a) shows the resulting $W_r$ for our fiducial parameters; the orange and cyan lines are the electron and positron beam density contour, respectively, at the $1/e^2$ density of the peak value. We note a negligible transverse force at the head of the beam ($\xi\gtrsim\SI{15}{\micro\meter}$); however, the force around the beam center and the tail ($\xi\lesssim\SI{10}{\micro\meter}$) is sufficiently strong that one may expect the beam to undergo significant dynamics. In particular, the force is focusing (defocusing) for positrons (electrons) near the axis, i.e., $r \le \SI{15}{\micro\meter}$, and vice versa for $r \ge \SI{15}{\micro\meter}$ (note this value can be calculated from $dR/dr = 0$); thus, the beams tend to separate spatially as electrons tend to focus around $r = \SI{15}{\micro\meter}$ and positrons around the axis. Figure \ref{fig:init_fields}(b) shows the focusing fields for our fiducial parameters obtained from two-dimensional cylindrical symmetric particle-in-cell simulations soon after the fireball beam enters the plasma (grid parameters and simulation details are presented in the Appendix). We observe a good agreement in both simulated field shape and amplitude when compared with the theoretical result in Fig. \ref{fig:init_fields}(a). While linear theory is not expected to capture the interaction of either positron or electron beam with the plasma because $n_b/n_0 = 1$, the effective fireball beam density, as shown in Fig. \ref{fig:beam_profile}, is about $\sim n_0/4$, closer to the validity of the linear approximation. This leads to the good agreement observed in Fig. \ref{fig:init_fields}.

\subsection{Nonlinear evolution toward a hollow beam driver}
\label{sec:nonlin}
\begin{figure*}[t]
    \centering
    \includegraphics[width=.99\linewidth]{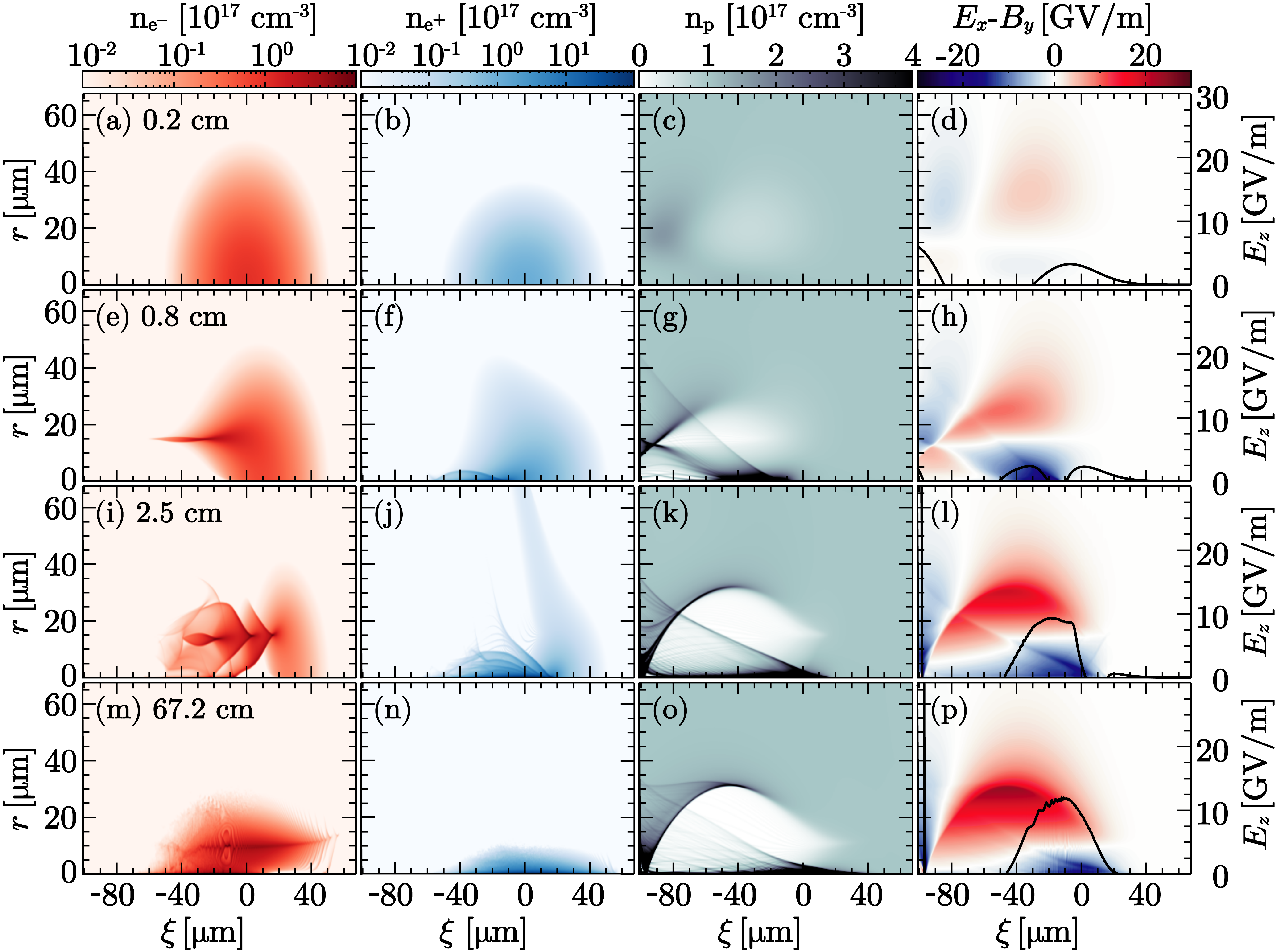}
    \caption{Nonlinear evolution of the fireball beam and the plasma. (a) Electron beam density, (b) positron beam density, (c) plasma electron density, and (d) focusing fields (colors) and longitudinal field at $r=0$ (line) after \SI{0.2}{\centi\meter} propagation in plasma. Panels (e) to (h), (i) to (l), and (m) to (p) show the same quantities after \SI{0.8}{\centi\meter}, \SI{2.5}{\centi\meter}, \SI{67.2}{\centi\meter} propagation in plasma, respectively.}
    \label{fig:beamevol}
\end{figure*}

We performed 2D cylindrical particle-in-cell simulations to follow the non-neutral fireball beam dynamics for long propagation distances (see Appendix for simulation details). We consider beams with \SI{10}{\giga eV}, no energy spread, and no initial emittance to focus on the fundamental processes governing the dynamics; the emittance role will be examined in Sec. \ref{sec:imp}(\ref{sec:emitt}). We have compared full 3D simulations with lower energy (\SI{1}{GeV}), enabling faster transverse dynamics, with the 2D cylindrical simulations, resulting in an excellent agreement, which makes us confident in the validity of the results presented henceforth. Figure \ref{fig:beamevol} shows the electron and positron beam densities, the plasma density, and the electromagnetic fields at several points over their propagation. The lineouts in the fourth column of Fig. \ref{fig:beamevol} are the longitudinal electric field at $r=0$; we only show positive values because our main interest will be identifying positron accelerating regions ($E_z>0$). 
As mentioned earlier, the wakefield driven by the non-neutral fireball tends to separate the electron and positron beam spatially; furthermore, the modifications in the beam profile drive even stronger plasma waves that reinforce the beam separation tendency. This feedback loop seems a manifestation of the current filamentation instability, which was one of the main motivations that led to the idea of studying fireball beam propagation in plasmas \cite{2015Sarri, 2018Shukla, 2020Shukla, 2021Arrowsmith}.

One can observe the initial beam dynamics by comparing the initial conditions in Figs. \ref{fig:beamevol}(a)-\ref{fig:beamevol}(b) with Figs. \ref{fig:beamevol}(e)-\ref{fig:beamevol}(f). Beam electrons accumulate near $r \approx \SI{15}{\micro\meter}$, and the positrons near the axis. The plasma wave evolves significantly [compare Fig. \ref{fig:beamevol}(c) and Fig. \ref{fig:beamevol}(g)], and the fields become stronger [compare Fig. \ref{fig:beamevol}(d) and Fig. \ref{fig:beamevol}(h)]. Plasma electrons form a dense filament on-axis, tending to neutralize the high density of positrons in the region. For larger radii, at positions where the electron beam focuses, we have regions completely void of plasma electrons, indicating that the plasma wave is strongly nonlinear. The plasma wave in Fig. \ref{fig:beamevol}(g) resembles the regime studied in Refs. \cite{2014Vieira,2015Jain} for hollow beams. 

Another step in the evolution of the beams is the development of the fishbone-like structures, shown in Figs. \ref{fig:beamevol}(i)-\ref{fig:beamevol}(j). The structures arise from betatron oscillations with different frequencies along the beam \cite{2015Jain}. These oscillations can be detrimental for hollow beams, as described in Ref. \cite{2019Jain}, which leads to the beam collapsing on-axis. Here, this effect seems to be mitigated by the presence of the positron beam near the axis.
However, we do observe the collapse of the electron beam driver under certain conditions [c.f.  Sec. \ref{sec:imp}(\ref{sec:beam_radii})]. We also note that some positrons escape radially because they are in a defocusing region for positively charged particles; this does not significantly impact the configuration. Figure \ref{fig:beamevol}(l) shows regions where accelerating gradients are on the order of $\SI{10}{\giga V/\meter}$.

Figures \ref{fig:beamevol}(m)-\ref{fig:beamevol}(p) show the system after $\SI{67}{\centi\meter}$ propagation in plasma. The configuration self-consistently evolved to be similar to the one studied in Ref. \cite{2015Jain}. As a result, a large fraction of the positrons were accelerated, detailed next.
\subsection{Positron acceleration}
\label{sec:pos}
\begin{figure}[t]
    \centering
    \includegraphics[width=.99\linewidth]{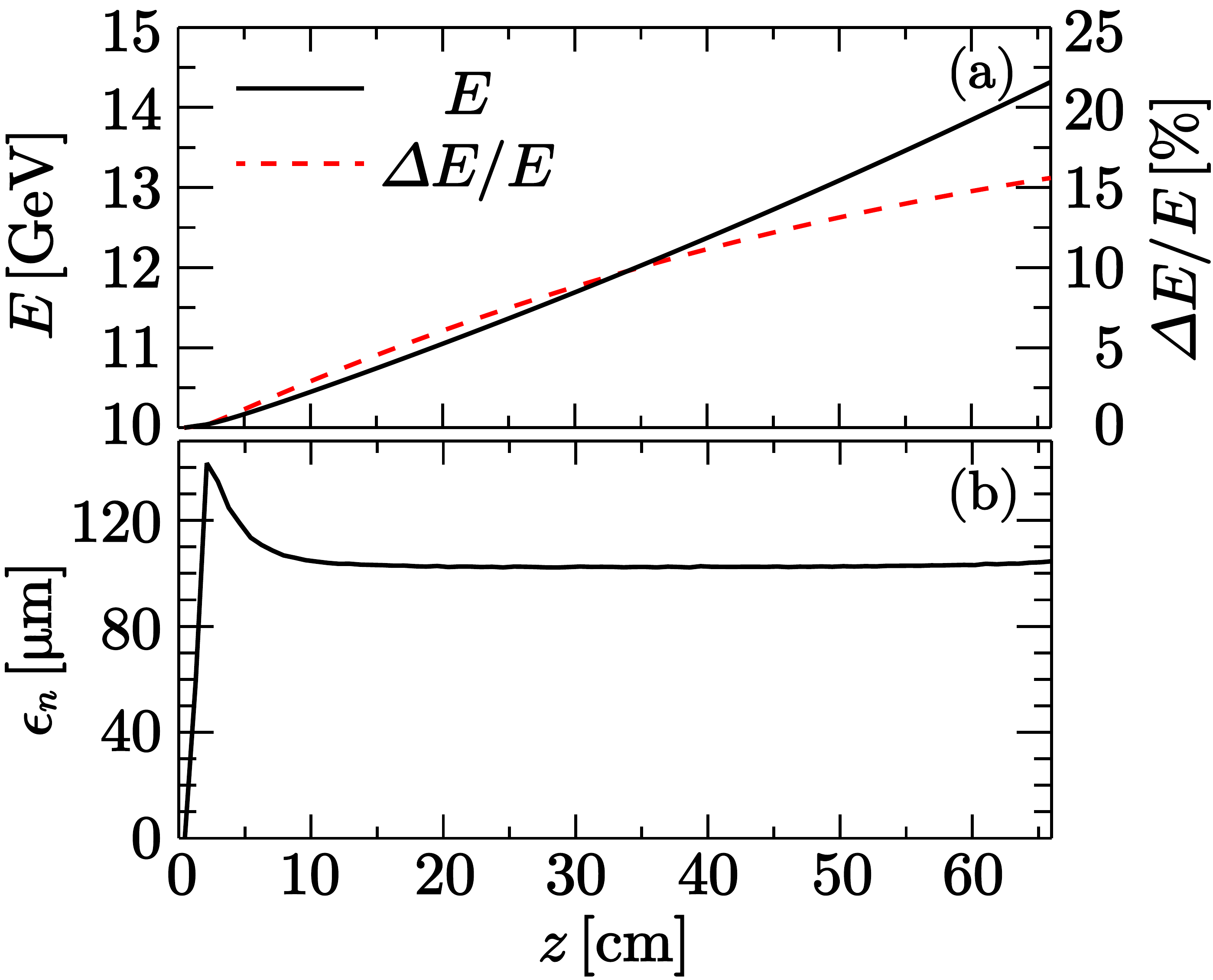}
    \caption{(a) Energy (black, solid line) and correlated energy spread (red, dashed line) positron beam evolution as it propagates in plasma. (b) Normalized emittance evolution.}
    \label{fig:beamparams}
\end{figure}

The non-neutral fireball beam self-consistently evolves toward a near-hollow electron beam circumscribing a positron beam. Thus, as described in Refs. \cite{2014Vieira,2015Jain}, one may expect the positron beam to accelerate. We confirm this by following the evolution of the positron beam in our fiducial simulation. The positron charge trapped in accelerating and focusing fields is $\SI{240}{\pico C}$, $40\%$ of the initial charge in the fireball beam. Figure \ref{fig:beamparams}(a) shows the average energy and the energy spread evolution over $\SI{67}{\centi\meter}$ propagation in plasma. Withing the beam, the accelerating gradient can be as large as $>\SI{10}{\giga V/\meter}$ [see. Fig. \ref{fig:beamevol}(p)], but on average, it is $\SI{6.6}{\giga V/\meter}$; this variation of the accelerating field along the beam leads to the RMS energy spread growth observed in Fig. \ref{fig:beamparams}(a). Figure \ref{fig:beamparams}(b) displays the normalized emittance 
\begin{equation}
    \epsilon_n = \frac{1}{mc}\left(\left<x^2\right>\left<p_x^2\right>-\left<xp_x\right>^2\right)^{1/2}
\end{equation}
evolution, which shows an appreciable growth during the nonlinear evolution described in Sec. \ref{sec:dyn}(\ref{sec:nonlin}). After \SI{10}{\centi\meter}, the emittance stabilizes, and it remains nearly constant during the rest of the propagation.
\section{Tolerance studies}
\label{sec:imp}
The acceleration of positrons using non-neutral fireball beams requires the spatial overlap of an electron and a positron beam. One should assume that the beams likely will present imperfections and variations from shot to shot for a possible experimental application. In this section, we perform a preliminary tolerance study that helps to evaluate the tolerances under different circumstances.
\subsection{Beam transverse size requirements}
\label{sec:beam_radii}

\begin{figure}[b]
    \centering
    \includegraphics[width=.99\linewidth]{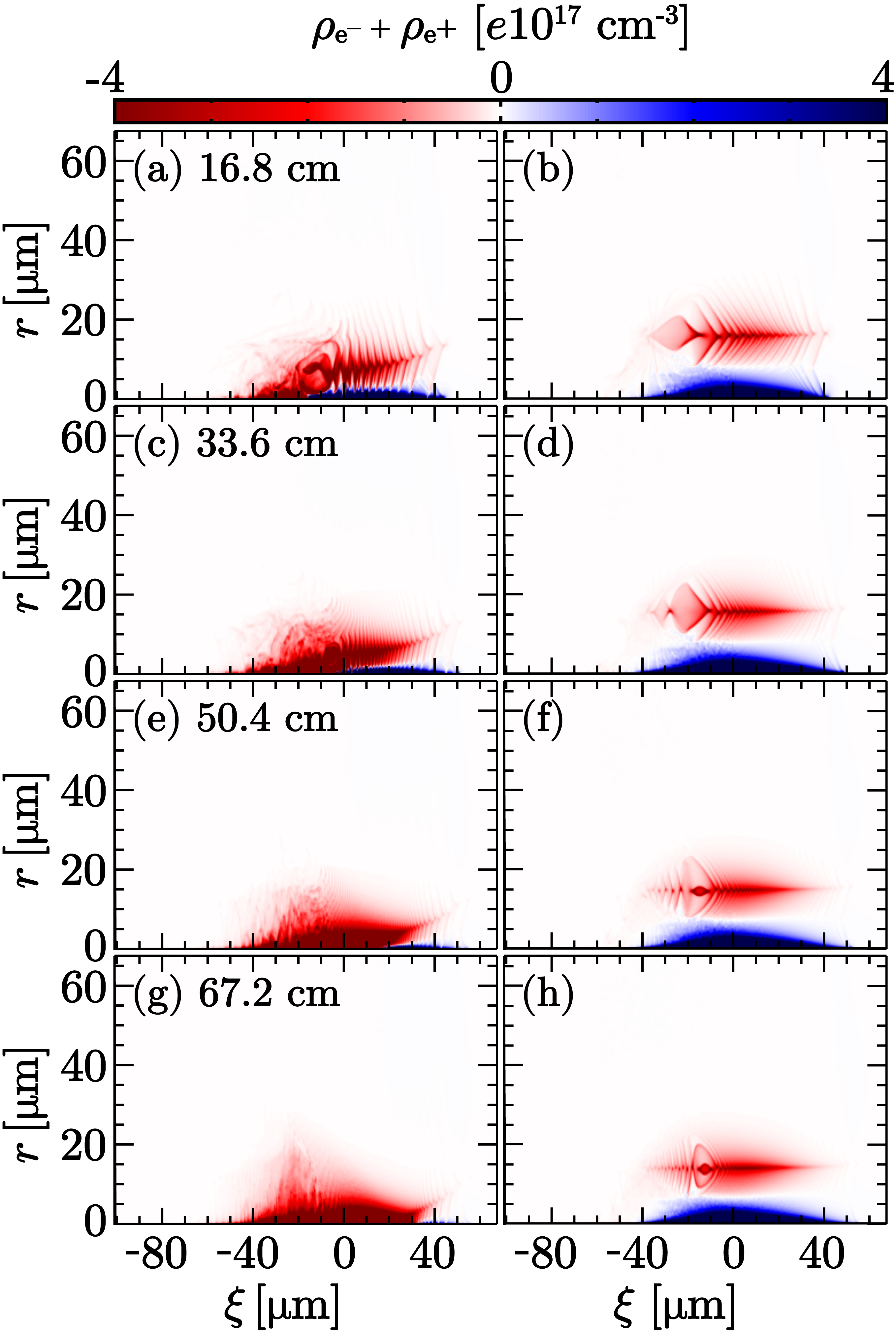}
    \caption{Fireball beam charge density [Eq. \eqref{eq:dens_hollow}] for $\sigma_{r,\mathrm{e}^+} = \SI{10}{\micro\meter}$ (left column) and $\sigma_{r,\mathrm{e}^+} = \SI{13}{\micro\meter}$ (right column) keeping the other parameters the same as the fiducial. Results after (a), (b) \SI{17}{\centi\meter}; (c), (d) \SI{33.6}{\centi\meter}; (e), (f) \SI{50.4}{\centi\meter}; and (g), (h) \SI{67.2}{\centi\meter} propagation in plasma.}
    \label{fig:rad}
\end{figure}
After an initial period when the fireball beam evolves into a hollow electron and a focused positron beam, the positron acceleration process becomes stable in our fiducial simulation until the electron bunch loses a significant part of its energy. However, the hollow electron beam may collapse on-axis earlier than energy depletion depending on the relation between the electron and positron beam transverse sizes. The collapse was first described in Ref. \cite{2019Jain} for a hollow electron beam propagating in plasma. The electron and the positron beam radii control the equilibrium position that the electron beam evolves toward [$W_r = 0$ in Fig. \ref{fig:init_fields}(a)]; if this position is sufficiently close to the axis, the electron beam may collapse on-axis after a few betatron oscillations \cite{2019Jain}. Nevertheless, we expect the collapse to be slightly different from the one described in Ref. \cite{2019Jain}, where the collapse occurs due to a significant accumulation of electrons near the axis as a result of the fishbone-like structures. Figure \ref{fig:beamevol}(i) shows that the fishbone structure reaches the axis in our example, but due to the presence of the positron beam, the collapse does not occur. 

According to our numerical simulations, the collapse of the electron beam driver can be prevented when the density in the fishbone structure is lower than the positron density near the axis. Figure \ref{fig:rad} displays two examples showing the total charge density of the fireball beam [Eq. \eqref{eq:dens_hollow}]; in the first (left column), the positron beam radius is $\sigma_{r,\mathrm{e}^+} = \SI{10}{\micro\meter}$; in the second (right column), $\sigma_{r,\mathrm{e}^+} = \SI{13}{\micro\meter}$. In the first example, we observe the differential shift of the electron beam toward the axis described in Ref. \cite{2019Jain}, which eventually leads to the collapse of the electron beam on-axis and defocusing of the positron beam. In the second example, the configuration is stable, with negligible changes over \SI{67.2}{\centi\meter} propagation in plasma.

Our results show that the collapse occurs if the positron beam transverse size is sufficiently smaller than that of the electron beam. Thus, the charge is also smaller (since we kept the peak density constant), and there are fewer positrons to neutralize the electrons oscillating in the fishbone structure. However, we note that the closer the positron bunch radius is to that of the electron driver, the lower the accelerating fields because the charge separation will not be as predominant. Therefore, we have a trade-off between the accelerating gradient and the long-time stability of the positron acceleration. Nevertheless, non-zero emittance may help avoid beam collapse under appropriate conditions, as it will be discussed in Sec. \ref{sec:imp}(\ref{sec:emitt}).
\subsection{Beam divergence role}
\label{sec:emitt}
The relation between the electron and positron beam divergences plays a significant role in having long-term stable acceleration. As shown in Fig. \ref{fig:init_fields}, the beam front propagates almost as in the vacuum since the focusing forces are negligible in that region, and this part of the beams will diverge with time. If the positron beam diverges slower than the electron, positron acceleration can happen over large propagation distances.

To demonstrate this, we consider the transverse envelope equation for each species $\alpha$. The transverse envelope ($\sigma_{r,\alpha}$) measures the RMS position of the beam particles, and its evolution is governed by \cite{2016Ferrario}
\begin{equation}
    \frac{d^2 \sigma_{r,\alpha}}{dz^2} - \frac{\theta_{\alpha}^2}{\sigma_{r,\alpha}} = 0,
    \label{eq:env}
\end{equation}
where 
\begin{equation*}
\theta_{\alpha} = \left<\frac{p_{r,\alpha}^2}{p_{z,\alpha}^2}\right>^{1/2},
\end{equation*}
is the RMS beam divergence. Equation \eqref{eq:env} describes the propagation of the beams in the absence of focusing forces, thus it is only valid here for the fireball beam front.  Electron and positron beams available in many accelerator facilities have emittance on the order of a few to tens of $\SI{}{\micro\meter}$ \cite{2014Tavares, 2016Aschikhin, 2019Yakimenko}. We use these values as a reference to calculate the initial divergence in the following analysis.
\begin{figure}[t]
    \centering
    \includegraphics[width=.99\linewidth]{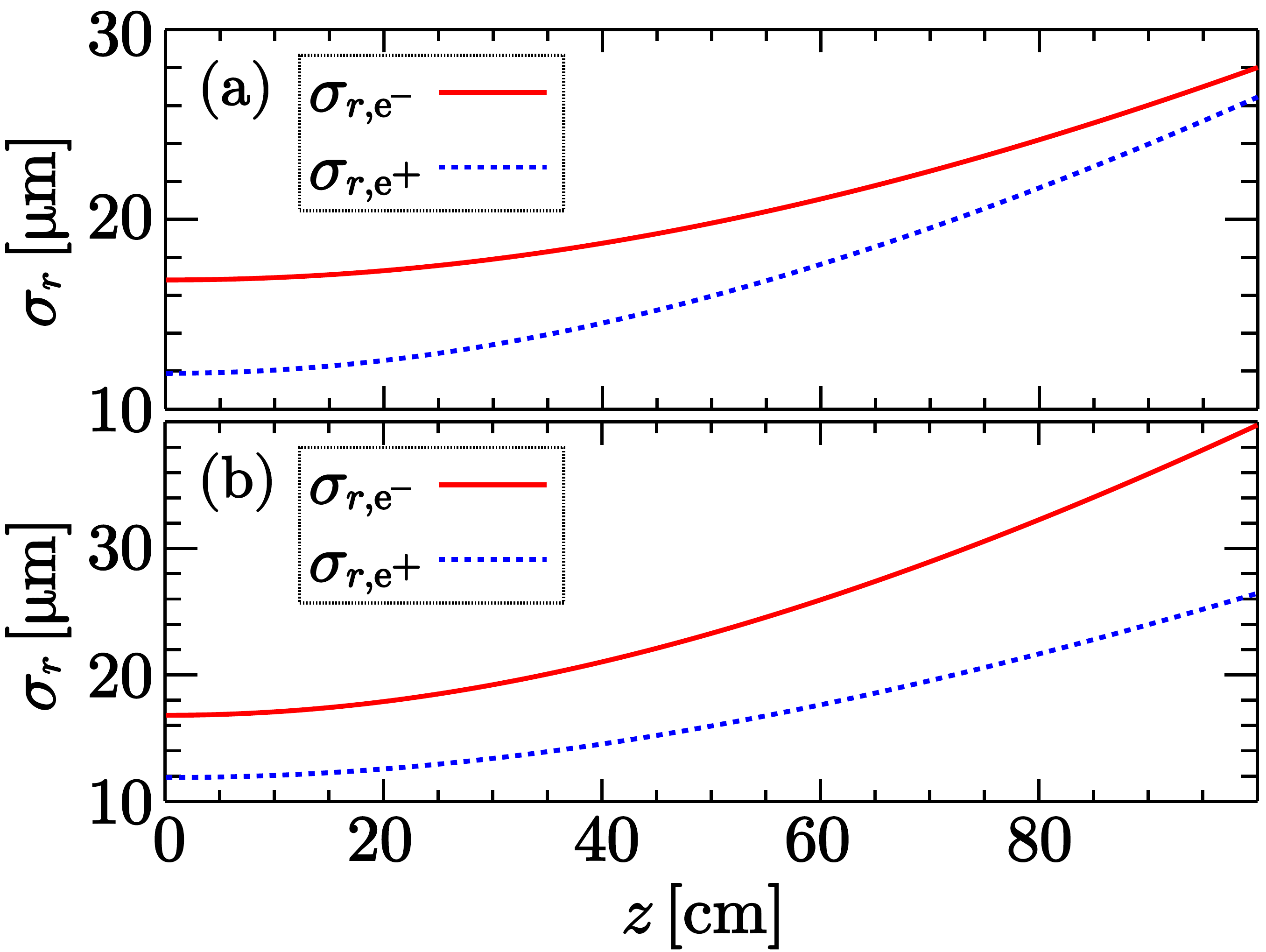}
    \caption{Electron and positron envelope evolution [Eq. \eqref{eq:env}]. (a) Both beams start with same transverse momentum spread $0.4\,m_ec$. (b) The electron beam has a larger momentum spread $0.6\,m_ec$ than the positron $0.4\,m_ec$. All the remaining parameters are the same as the fiducial.}
    \label{fig:env_div}
\end{figure}

Even though $\sigma_{r,{\mathrm{e}^-}}(z=0) > \sigma_{r,{\mathrm{e}^+}}(z=0)$, if the positron beam diverges faster than the electron, the positron envelope may eventually be larger than the electron. While the fireball beam front does not play a significant role in the positron acceleration, it does seed the process that modifies the electron beam profile from a Gaussian to a hollow beam. Figure \ref{fig:env_div} shows two examples of solutions of Eq. \eqref{eq:env} for our fiducial parameters but considering distinct divergences. In Fig. \ref{fig:env_div}(a), the positron and electron beam divergences are the same, with $\big<p_{r,\mathrm{e}^+}^2\big> = \big<p_{r,\mathrm{e}^-}^2\big> = 0.4\,m_ec$, resulting in initial emittances of $\epsilon_{n,\mathrm{e}^+} = \SI{4.7}{\micro\meter}$ and $\epsilon_{n,\mathrm{e}^-} = \SI{6.7}{\micro\meter}$. We notice that the positron envelope tends closer to the electron, and it will eventually surpass it. In Fig. \ref{fig:env_div}(b), we use a larger value for $\big<p_{r,\mathrm{e}^-}^2\big> = 0.6\,m_ec$, resulting in $\epsilon_{n,\mathrm{e}^-} = \SI{10}{\micro\meter}$. In this case, the electron envelope is always larger than the positron.

\begin{figure}[t]
    \centering
    \includegraphics[width=.99\linewidth]{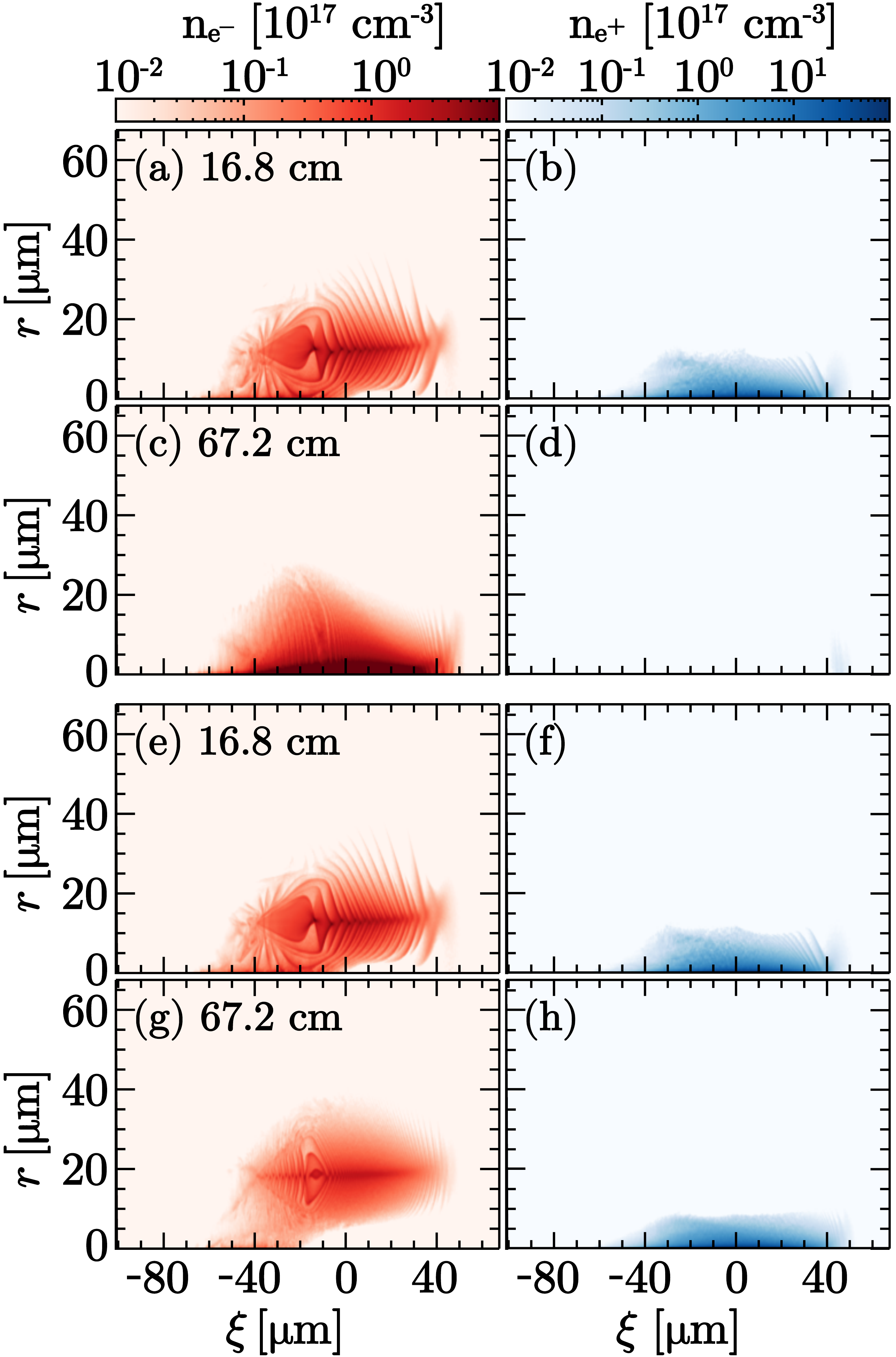}
    \caption{Panels (a) to (d): electron (left column) and positron (right column) density for the initial conditions used in Fig. \ref{fig:env_div}(a). Panels (a) and (b) are after \SI{16.8}{\centi\meter} propagation in plasma, while (c) and (d) after \SI{67.2}{\centi\meter}. Panels (e) to (h) present equivalent results for the initial conditions used in Fig. \ref{fig:env_div}(b).}
    \label{fig:div}
\end{figure}
We performed particle-in-cell simulations considering the conditions above for the beam divergence. Figures \ref{fig:div}(a)-\ref{fig:div}(d) show results for the first example [the same parameter as in Fig. \ref{fig:env_div}(a)] and Figs. \ref{fig:div}(e)-\ref{fig:div}(h) for the second [same as Fig. \ref{fig:env_div}(b)]. We first focus on early propagation, i.e., \SI{16.8}{\centi\meter} in plasma. The comparison of Figs. \ref{fig:div}(a) and (e) and Figs. \ref{fig:div}(b) and (f) show nearly identical results, which we argue are due to negligible changes to the envelopes at that point [note that all the curves in Fig. \ref{fig:env_div} are approximately at the same value as their initial condition at $z=\SI{16.8}{\centi\meter}$]. Nevertheless, the long-term propagation is considerably different: in the first example, as the front of the positron beam envelope increases faster than the electron, the plasma waves driven become less suitable for positron focusing, leading to the collapse of the hollow electron beam on-axis [Figs. \ref{fig:div}(c)] and the defocusing of the positrons until nearly all the positrons escape radially [Figs. \ref{fig:div}(d)]; in the second example, the hollow electron and the positron beam remain with their appropriate shape for positron acceleration after \SI{67.2}{\centi\meter} propagation in plasma [Figs. \ref{fig:div}(g)-(h)].
These results indicate that Eq. \eqref{eq:env} can estimate if the formation of the hollow electron beam and the long-term positron propagation are stable due to non-zero emittance.

\subsection{Time jitter}
\begin{figure}[t]
    \centering
    \includegraphics[width=.99\linewidth]{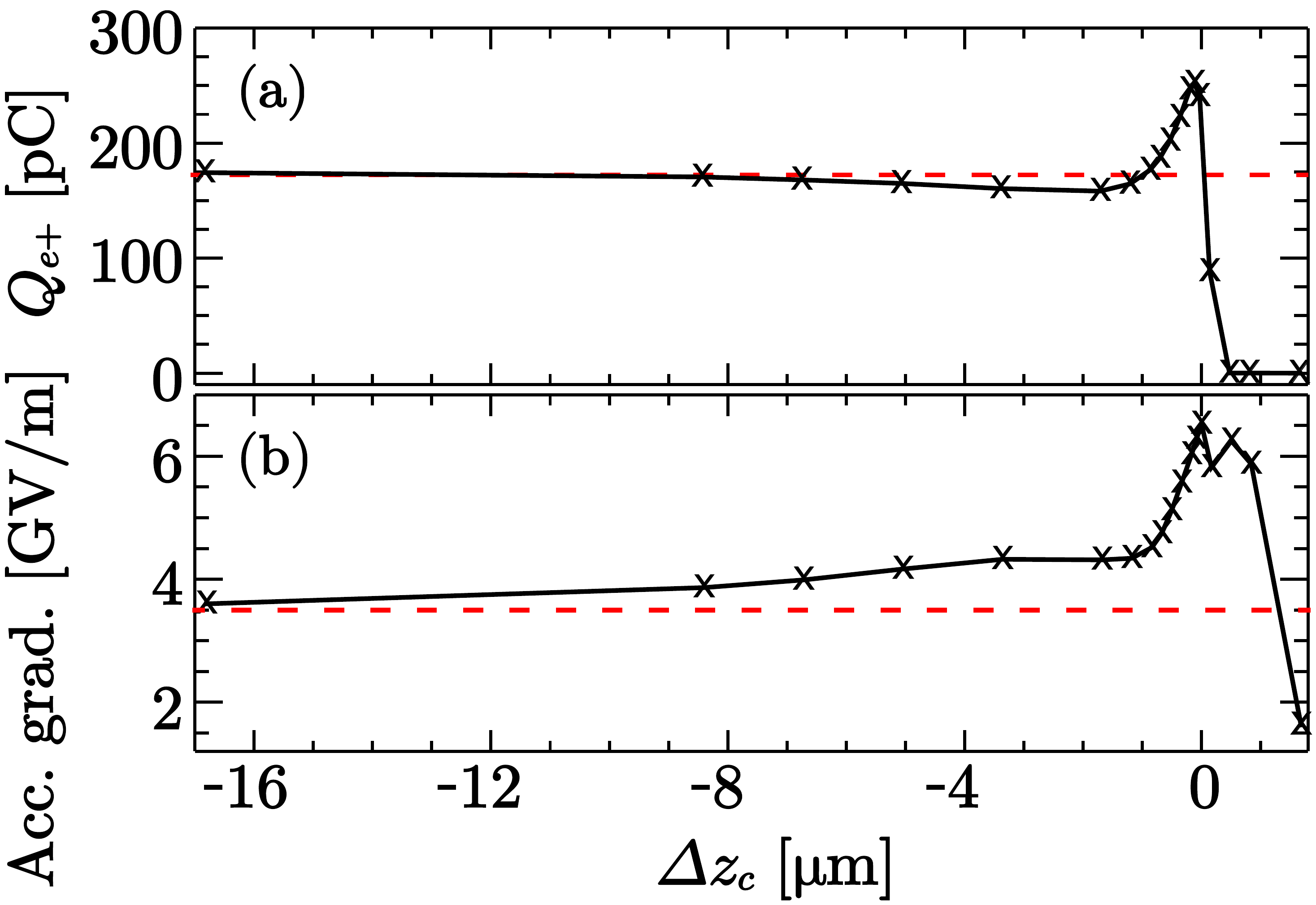}
    \caption{(a) Positron beam accelerated charge after \SI{67.2}{\centi\meter} propagation in plasma as a function of the initial time jitter. The symbols represent simulation results. For comparison, the red, dashed line shows the value calculated in the absence of the electron beam, i.e., the self-loaded regime. (b) Equivalent results for the accelerating gradient measured in simulations.}
    \label{fig:jitter}
\end{figure}
To understand how the time jitter may affect the scheme, we define $\Delta z_c = z_{c,\mathrm{e}^-}(t=0) - z_{c,\mathrm{e}^+}(t=0)$ as the electron and positron beam center initial jitter. Note that if $\Delta z_c>0$ ($\Delta z_c<0$), the electron (positron) beam is ahead of the positron (electron). If $\Delta z_c\gg 0$, the system becomes equivalent to a positron beam in the wake of an electron beam in a quasi-linear or nonlinear regime, which is predominantly defocusing for positrons. On the other hand,  if $\Delta z_c\ll 0$, the positron beam is driving the wakefield, and one may accelerate a fraction of the positrons in the self-loaded wakefield \cite{2015Corde}.

In between these two limits, we have the configuration studied thus far when $\Delta z_c = 0$. We performed several simulations changing $\Delta z_c$ (the other parameters are the same as our fiducial case) to grasp the transition between the different regimes. Figure \ref{fig:jitter} displays the accelerated positron charge and the average accelerating gradient after $\SI{67}{\centi\meter}$ propagation in plasma. For comparison, the red dashed lines are the values obtained in the absence of the electron beam, i.e., in the self-loaded wakefield regime. We note in Fig. \ref{fig:jitter}(a) that the charge steeply drops to zero when $\Delta z_c > 0$; when $\Delta z_c < 0$, there is a smoother transition between the maximum charge (for $\Delta z_c \approx 0$) to values close to the self-loaded regime. Figure \ref{fig:jitter}(b) reveals that the accelerating gradient is also maximum for $\Delta z_c \approx 0$, and the value also changes rather quickly to the self-loaded regime when $\Delta z_c < 0$ (on the order of $\Delta z_c \approx \SI{-1}{\micro\meter}$).
\subsection{Transverse misalignment}
\label{sec:mis}
When the beams are aligned, they drive a symmetric mode that leads to the nonlinear evolution toward a hollow-shaped electron beam. However, when the beams are misaligned, we found that they tend to tilt in opposite directions when one reaches the nonlinear regime. In that case, the waves driven can be decelerating for most positrons. 

While this seems prohibitive, we found parameters that enabled positron acceleration over shorter distances.
\begin{figure}[t]
    \centering
    \includegraphics[width=.99\linewidth]{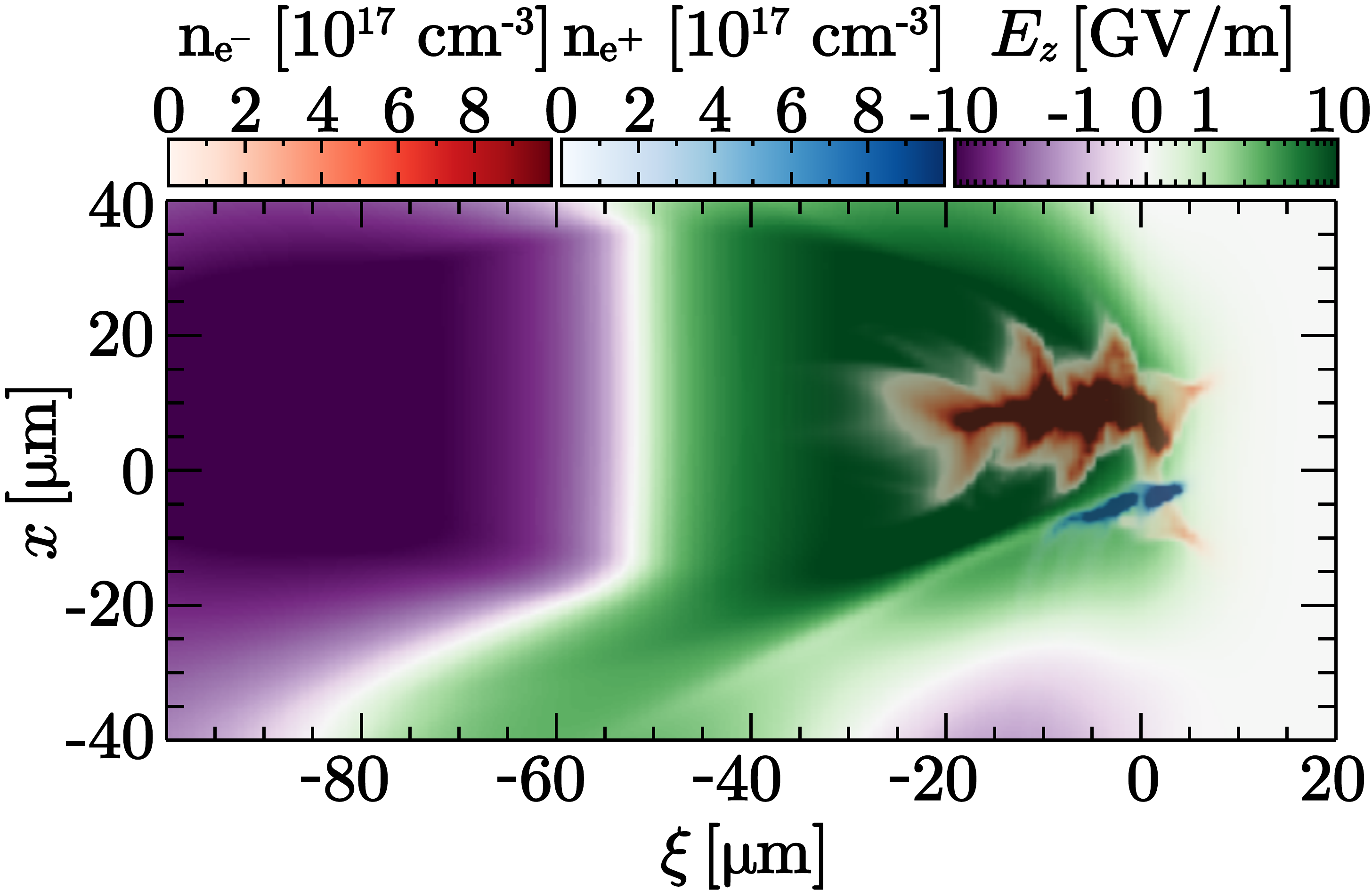}
    \caption{Electron beam, positron beam, and longitudinal fields after \SI{16}{\milli\meter} propagation in plasma for a case in which the beams had an initial misalignment.}
    \label{fig:misal}
\end{figure}
Simulations showed that the effect of transverse misalignments is less important for shorter beams, as the beam tilting becomes less predominant. Also, wider transverse density profiles for the electron beam (e.g., super-Gaussian instead of Gaussian) can help mitigate the effects of the misalignment. Figure \ref{fig:misal} shows the result of one three-dimensional simulation in which the positron beam is still accelerating after significant nonlinear evolution. In this simulation, both beams start with \SI{1}{\giga eV}, thus reducing the betatron period to have faster transverse dynamics. The longitudinal profile of both beams is Gaussian  with $\sigma_z=\SI{8.4}{\micro \meter}$, half of the value used in the fiducial simulations. The positron beam still has a Gaussian transverse profile with $\sigma_{r,\mathrm{e}^+} = \SI{15.1}{\micro \meter}$; the electron beam has a super-Gaussian profile $\rho\propto \exp(-r^4/\sigma_{r,\mathrm{e}^-}^4)$ with $\sigma_{r,\mathrm{e}^-} = \SI{25.2}{\micro \meter}$. The initial beam misalignment is $\SI{0.84}{\micro \meter}$.

As shown in Fig. \ref{fig:misal}, the electron beam still resembles a doughnut shape at $\xi\approx 0$, and the positron beam is compressed in a region of accelerating fields. At this time, \SI{370}{} of the initial \SI{490}{\pico C} positron beam charge is in accelerating fields, with an average accelerating gradient of $\SI{1}{\giga V/m}$ over  \SI{16}{\milli\meter} propagation in plasma. However, the electron beam presents a strong compression for $x>0$ [Fig. \ref{fig:misal}], and drives strong wakefields. Thus, the regime is not as efficient as the symmetric mode since the electron beam loses energy at a rate almost one order of magnitude higher than the positron gains. Additionally, the long-term stability of the system is not guaranteed as the beams are still evolving at that point.

An initial misalignment proved to be the most challenging imperfection to overcome, particularly when compared with other schemes \cite{2021Silva, 2021Zhou, 2020Diederichs,2022Diederichs}. As the beam profile evolution occurs due to an unstable feedback loop, minor misalignments can seed a different mode that will be substantially different when reaching the nonlinear regime. As these three-dimensional simulations are costly, we could not thoroughly explore many ideas, such as beams with asymmetries, different energies, tailored profiles, or the use of external magnetic fields, which we intend to do in future work. Further numerical studies will be required if one wishes to demonstrate robustness. The analogous scheme using a laser beam, described next in Sec. \ref{sec:laser}, could also be less sensitive to a misalignment; however, it would be even more computationally demanding given the need to resolve the laser wavelength and the initial misalignment.

\section{Laser beam}
\label{sec:laser}
Thus far, we focused on a configuration with an electron and a positron beam that self-consistently evolved into a hollow electron beam and a focused positron beam on-axis. We now devote our attention to an analogous scheme, replacing the electron with a laser beam. Since the laser ponderomotive force tends to repel the background plasma electrons, the positron beam must be sufficiently dense ($n_{e+}\gtrsim n_p$) to counteract this effect and attract the electrons to the axis to form a high-density filament. In addition, the ponderomotive force should dominate for large radii to create an electron sheath away from the axis. When these conditions are met, the laser beam will tend to self-focus away from the symmetry axis, evolving toward a hollow beam.

\begin{figure}[t]
    \centering
    \includegraphics[width=.99\linewidth]{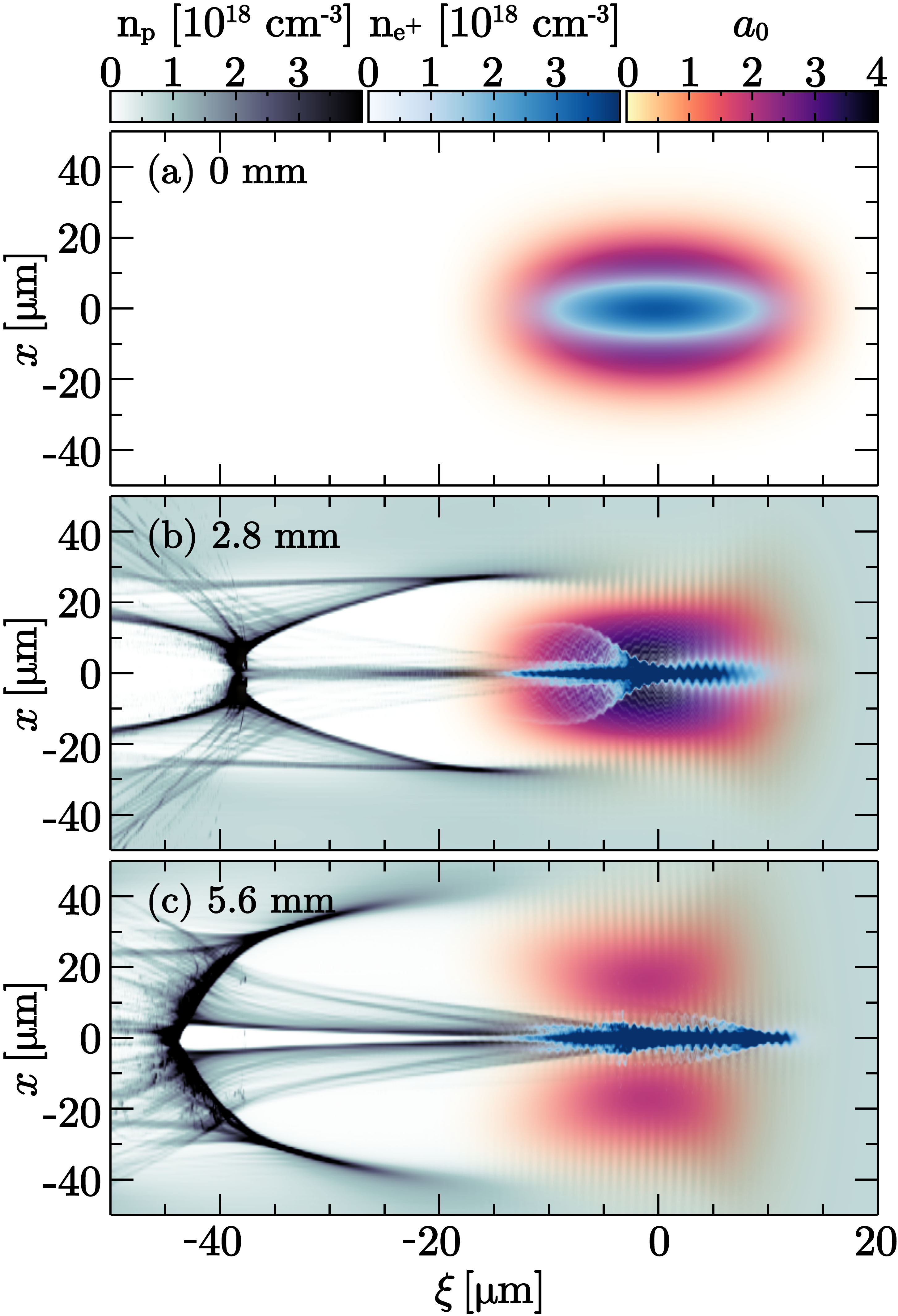}
    \caption{Plasma density, positron density, and laser envelope for simulations overlapping a positron beam with a laser beam. Panel (a) is the initial condition. Panels (b) and (c) are after \SI{2.8}{\milli\meter} and \SI{5.6}{\milli\meter} propagation in plasma, respectively.}
    \label{fig:laserevol}
\end{figure}

Figure \ref{fig:laserevol} shows three-dimensional particle-in-cell simulation results that corroborate the evolution of the positron and laser beams. We consider a laser with $\lambda =\SI{800}{\nano\meter}$ wavelength, $\tau=\SI{100}{fs}$ pulse duration, $a_0=4$ normalized vector potential, and $w_0=\SI{22}{\micro\meter}$ beam waist on the focal spot at the plasma entrance. The positron beam has \SI{5}{\giga eV} energy, \SI{2}{\nano C} charge, $\sigma_{r,\mathrm{e}^+} =\SI{6}{\micro\meter}$, and $\sigma_{z,\mathrm{e}^+} =\SI{22}{\micro\meter}$, and the plasma has a uniform profile with a density of $\SI{8.6e17}{\centi\meter^{-3}}$. Figure \ref{fig:laserevol}(a) displays the initial condition with the positron and laser (represented by its envelope) beams overlapping before entering the plasma. After a few \SI{}{\milli\meter} propagation [Fig. \ref{fig:laserevol}(b)], we note a strong compression of the positron beam and the formation of an electron filament on-axis. After further propagation, as shown in Fig. \ref{fig:laserevol}(c), the laser's peak intensity moves away from the axis and forms a doughnut-like shape. The intensity at this point is lower than at the start due to the beam's divergence; this effect could be mitigated using a parabolic density profile, like in standard laser-plasma accelerators \cite{2019Gonsalves}. In this example, the average accelerating gradient after $\SI{5.6}{\centi\meter}$ of propagation in plasma is $\SI{50}{\giga V/m}$, and 47\% ($\approx\SI{950}{\pico C}$) of the initial positron charge is accelerated. The emittance and energy spread grow to around $\epsilon_{x,y}\approx\SI{80}{\micro\meter}$ and $3\%$, respectively.

\section{Discussion}
\label{sec:disc}
Plasmas accelerators could be a high-gradient alternative to linear electron-positron colliders. As argued earlier, plasma-based acceleration is much harder for positrons than for electrons. In our work, we show that the scheme can accelerate positrons; however, due to the nonlinear beam evolution, the beam quality is degraded. Future developments for the scheme will focus on ideas for better stability and beam quality. One of the main advantages of the scheme is that it was able to accelerate a significant amount of charge over large distances, something that is not trivial in other schemes that are more fragile to beam loading.

Another outstanding question regards staging, which is likely required to achieve the energy goal for colliders. The positron beam is highly focused in both the beam [Sec. \ref{sec:dyn} (\ref{sec:nonlin})] and optically [Sec. \ref{sec:laser}] driven schemes. It is an open question how these focused beams would interact with similar drivers in another stage. This question will be the subject of future work. Yet, we argue that even a single stage could be beneficial to accelerate beams to significant energies before using a different acceleration scheme (e.g., \cite{2019Diederichs, 2021Silva}), as long as there are ways to mitigate the emittance degradation previously observed.

Our results may also be relevant in the context of current filamentation for laboratory or astrophysical plasmas. A neutral fireball beam propagation in plasma can lead to the generation of small-scale filaments, up to the plasma skin depth \cite{2020Shukla}. We have shown that imperfections such as charge unbalance or misalignments may result in qualitatively different regimes where filamentation has long-scale filaments on the order of the beam size.

\section{Conclusions}
\label{sec:concl}
In this work, we have proposed a scheme for plasma-based positron acceleration. We have shown that, under appropriate conditions, an overlap of a positron with an electron or laser beam can self-consistently evolve toward conditions proper for positron acceleration, namely hollow electrons or laser beams with positrons focused on-axis. This seems to be a manifestation of the current filamentation instability.

A key factor here is that the initial shape of the beams is Gaussian, something more readily available in laboratories when compared to starting with a hollow electron or laser beam. One drawback is that both beams undergo significant changes as there is a feedback loop between the beam dynamics and the plasma response; therefore, the beams can lose quality during their evolution, and the scheme can have some stringent tolerances. 

We have predicted conditions for stability under certain imperfections. However, transverse misalignments proved to be challenging to overcome, and future work will focus on ideas to demonstrate if the scheme can be robust. Additionally, the scheme with a laser instead of the electron beam requires a more systematic study since it is a configuration that might be less sensitive to imperfections.


\begin{acknowledgments}
We gratefully acknowledge PRACE for awarding us access to MareNostrum at BSC (Spain) and EuroHPC for awarding us access to LUMI-C at CSC (Finland). This work was supported by EU Horizon 2020 R\&I Grant No. 653782 (EuPRAXIA) and No. 871124 (Laserlab-Europe), EU Horizon Europe R\&I Grant No. 101079773 (EuPRAXIA-PP), and FCT (Portugal) Grant No. SFRH/IF/01635/2015.
\end{acknowledgments}
\appendix*
\section{Particle-In-Cell Simulations Details}
All simulations were performed using the particle-in-cell (PIC) code OSIRIS \cite{2002Fonseca,2013Fonseca}. We use a custom-built electromagnetic field solver to mitigate the numerical Cherenkov instability \cite{2017Li,2021Li}. Except for Sec. \ref{sec:imp}(\ref{sec:mis}) and Sec. \ref{sec:laser}, all simulations shown are two-dimensional with cylindrical symmetry. The grid comprises square cells with \SI{0.33}{\micro\meter} sides; the electron beam, the positron beam, and the plasma electrons start with 64, 64, and 16 particles per cell, respectively. The PIC loop time-step is \SI{0.56}{fs}. The simulation shown in Sec. \ref{sec:imp}(\ref{sec:mis}) is three-dimensional with cubic cells with \SI{0.84}{\micro\meter} with edges; all species start with 8 particles-per-cell, and PIC time-step is \SI{1.40}{fs}. Finally, the simulation shown in Sec. \ref{sec:laser} is three-dimensional and requires a higher longitudinal resolution to resolve the laser wavelength. The cells are edges are $\SI{38}{\nano\meter} \times \SI{0.72}{\micro\meter} \times \SI{0.72}{\micro\meter}$; all species start with 4 particles-per-cell, and PIC time-step is \SI{1.40}{fs}. 

\providecommand{\noopsort}[1]{}\providecommand{\singleletter}[1]{#1}%

\newpage

\end{document}